# Synthesis and properties of $CoO_2$, the $x = 0$ end member of the $Li_xCoO_2$ and $Na_xCoO_2$ systems

T. Motohashi[1], Y. Katsumata[1,2], T. Ono[1,2], R. Kanno[2], M. Karppinen[1,3], and H. Yamauchi[1,2]

[1]*Materials and Structures Laboratory, Tokyo Institute of Technology, Yokohama 226-8503, Japan*
[2]*Interdisciplinary Graduate School of Science and Engineering, Tokyo Institute of Technology, Yokohama 226-8502, Japan*
[3]*Laboratory of Inorganic and Analytical Chemistry, Helsinki University of Technology, P.O. Box 6100, FI-02015 TKK, Finland*

Abstract

We report here the synthesis of single-phase bulk samples of $CoO_2$, the $x = 0$ end member of the $A_xCoO_2$ systems ($A$ = Li, Na), from a pristine $LiCoO_2$ sample using an electrochemical technique to completely de-intercalate lithium. Thus, synthesized $CoO_2$ samples were found to be oxygen-stoichiometric and possess a crystal structure consisting of stacked triangular-lattice $CoO_2$ layers only. The magnetic susceptibility of the $CoO_2$ sample was revealed to be relatively large in its initial value and then level off as the temperature increases, suggesting that $CoO_2$ is a Pauli-paramagnetic metal with itinerant electrons.

1. Introduction

The layered cobalt oxide system, $Na_xCoO_2$, has attracted increased interests due to a variety of unconventional transport and magnetic properties. About $x = 0.7$, $Na_xCoO_2$ exhibits an unusually large thermoelectric power and a metallic conductivity [1]. It was revealed that, when $x$ is increased only by 0.05, *i.e.* at $x = 0.75$, thermoelectric power is significantly enhanced [2,3] and a spin-density-wave state appears below $T_m = 22$ K [4]. In the lower Na-content region, the $x \approx 0.35$ member readily absorbs water and the water derivative $Na_{0.35}CoO_2 \cdot yH_3O \cdot y'H_2O$ becomes a superconductor (with $T_c = 4.5$ K) [5]. It was also found [6] that the magnetism of the $Na_xCoO_2$ system is of the Curie-Weiss type for $x > 0.5$, while paramagnetic for $x < 0.5$, and for $x = 0.5$ electrical conduction is poor due to the appearance of a charge-ordered state.

The variety in the property with respect to the Na content ($x$) in the $Na_xCoO_2$ system is likely stemmed from some involved spin interactions. It was reported that the $x = 1.0$ end member, *i.e.* $NaCoO_2$, is a non-magnetic insulator containing low-spin ($S = 0$) $Co^{III}$ [7]. As $x$ decreases, the average valence of cobalt should gradually increase toward $+4$ such that the concentration of magnetic $Co^{IV}$ species may increase. In the $Na_xCoO_2$ system, antiferromagnetic ordering is suppressed since the cobalt atoms form a triangular lattice with geometrical frustration that may cause intriguing electronic/magnetic properties. Thus one may anticipate that the other end member, *i.e.* $CoO_2$, would form an ideal magnetically-frustrated system in which every site in the triangular lattice is occupied by magnetic $Co^{IV}$. Recent theoretical works based on (LDA+$U$) band calculations suggested [8,9] that the $CoO_2$ crystal is either a charge-transfer insulator or a metal with itinerant electrons, depending on the strength of on-site Coulomb interaction $U$. Bearing in mind that $CoO_2$ may be considered as the parent compound of $A_xCoO_2$ ($A$ = Li, Na), synthesis of $CoO_2$ is a first step for understanding the electronic structure of the $A_xCoO_2$ system.

To materialize oxides containing high-valent cobalt, employment of strongly oxidizing agents/conditions is indispensable. The oxidation of $Na_xCoO_2$ ($x$ = 0.7) by means of chemical reagents including $Br_2$ and $NO_2BF_4$ resulted in an incompletely Na-extracted form, *i.e.* $Na_{0.15}CoO_2$ [10]. Here to obtain pure $CoO_2$ we have employed an electrochemical oxidation technique for $LiCoO_2$, *i.e.* electrochemical de-intercalation of Li from $LiCoO_2$. Electrochemical techniques are believed to be among the most powerful ones for realizing high-valent transition-metal oxides. Since both $Na_xCoO_2$ and $LiCoO_2$ contain the $CdI_2$-type $CoO_2$ layers in common, the $x$ = 0 end members of the two systems, *i.e.* $Na_0CoO_2$ and $Li_0CoO_2$, are considered to be identical [11]. Note that $LiCoO_2$ is one of representative cathode materials for the lithium-ion secondary battery since the capability of electrochemical de-intercalation of Li is large enough.

Synthesis of $CoO_2$ samples through electrochemical oxidation was previously reported by Amatucci *et al.* [12] Their *in-situ* x-ray diffraction analysis of a $Li_xCoO_2$ cathode that had been placed in a dry plastic cell while being electrochemically charged revealed that Li ions were completely extracted and thus pure $CoO_2$ was obtained. Nonetheless, no physical property data of $CoO_2$ were reported. Therefore, it is highly desirable to obtain bulk samples of pure $CoO_2$ for the study of its physical properties. Here we report synthesis and some properties of pure $CoO_2$ bulk as materialized through electrochemical de-intercalation of Li from a LiCoO$_2$ bulk. Approximately 50 mg of CoO$_2$ bulk allowed us to successfully measure its dc-magnetic susceptibility with reliable accuracy.

2. Experimental

First stoichiometric LiCoO$_2$ samples were synthesized by a conventional solid-state reaction technique [13]. A mixture of Li$_2$CO$_3$ (Seimi Chemical Co. Ltd., 99.9%; dried at 120°C) and Co$_3$O$_4$ (Rare Metallic Co. Ltd., 99.9%; decarbonated at 450°C for 12 h in O$_2$) powders with the ratio of Li : Co = 1 : 1 was calcined at 600°C for 12 h in flowing O$_2$ gas. The calcined powder was ground, pelletized and fired at 900°C for 36 h in flowing O$_2$ gas with intermediate grinding. Results of x-ray powder diffraction (XRPD; Rigaku RINT-2000V equipped with a rotating anode; Cu $K_\alpha$ radiation) indicated that the pristine LiCoO$_2$ sample was of single phase with a rhombohedral structure.

Then electrochemical oxidation was carried out with a constant current (*i.e.* galvanometric) setup utilizing an air-tight flat cell made of stainless steel. An as-synthesized LiCoO$_2$ bulk pellet of diameter and thickness of ~10 mm and ~0.5 mm was set as the working electrode. No auxiliary agents (*e.g.* acetylene black and Teflon powder) were added to the bulk pellet to avoid any magnetic noise sources. The counter electrode was an aluminum metal disk with diameter of 15 mm. Employed was non-aqueous electrolyte consisting of 1 M LiPF$_6$ in 7 : 3 mixture solution of ethylene carbonate (EC) and diethyl carbonate (DEC) (of battery grade provided by Mitsubishi

Chemical Corporation). The cell was constructed in an argon-filled glovebox. The content of lithium in the resultant $CoO_2$ pellet was determined by means of inductively coupled plasma atomic-emission spectroscopy (ICP-AES; Shimadzu: ICPS-8100).

Since high-valent metal oxides tend to experience chemical instability when exposed to atmospheric moisture, sample handling and characterization were made such that the samples might be exposed to moisture as little as possible. After the electrochemical procedure, the sample was washed with anhydrous acetonitrile or dimethyl carbonate in an argon-filled glovebox and encapsulated into a Pyrex ampoule. XRPD analysis was carried out for the electrochemically treated sample which was set in an airtight sample holder filled with Ar gas. The oxygen content of the sample was determined based on hydrogen reduction experiments carried out in 5% $H_2$/Ar gas flow with a thermobalance (Pyris 1; Perkin Elmer). Magnetic susceptibility measurements were performed with a SQUID magnetometer (MPMS-XL; Quantum Design). An as-encapsulated sample was put in a cryostat for magnetic measurements: the influence of the ampoule on the measurement was calibrated accordingly.

3. Results and discussion

Single-phase bulk samples of the $CoO_2$ phase were successfully obtained through the aforementioned electrochemical de-intercalation of lithium from a stoichiometric $LiCoO_2$ bulk pellet. Figure 1(a) shows a typical voltage *vs* *x* plot for the $Li_xCoO_2$/Al-metal cell. In this experiment, a bulk of 50 mg of pristine $LiCoO_2$ was charged with a constant current of 0.1 – 0.2 mA (= 0.13 – 0.25 mA/cm$^2$) for 70 – 140 hours. The $x$ values were estimated through theoretical calculations based on Faraday's law with an assumption that the full amount of electricity due to the current was used for the electrochemical de-intercalation of Li from Li$_x$CoO$_2$. The cell voltage ($V$) gradually increased with elapsed time, *i.e.* with decreasing $x$ in Li$_x$CoO$_2$, and finally reached +4.78 V at $x$ = 0.0. As $x$ decreased below 0.1, the cell voltage increased rapidly to indicate the completion of lithium extraction. This feature is commonly seen in the de-intercalation of *e.g.* LiMn$_2$O$_4$ [14]. The *V-vs-x* curve is in good agreement with that previously reported by Amatucci *et al.* [12]. The only difference between the two curves is the voltage value at $x$ = 0.0, which is due to the different anode materials. The actual Li content ($x$) of the resultant CoO$_2$ sample was determined to be below the detection limit of the ICP-AES apparatus, *i.e.* smaller than 0.01. This ensures that the resultant CoO$_2$ samples are indeed the $x$ = 0 end member of the Li$_x$CoO$_2$ system.

To understand the phase evolution process during the course of Li de-intercalation from LiCoO$_2$, the derivative of the potential curve, *i.e. dV/dx*, is plotted as a function of Li content ($x$) in Fig. 1(b). A characteristic "dip-peak-dip" structure is clearly seen about $x$ = 0.5. Similar features were reported in previous works [12,15], attributing the structure to the appearance of a Li-ion ordered phase at $x$ = 0.5 as predicted by a first-principle calculation [16]. This marking for $x$ = 0.5 supports the accuracy of the $x$ value as calculated from Faraday's law. Furthermore, two broad peaks are seen at $x \approx$ 0.33 and 0.15 in the *dV/dx vs x* plot, which would indicate the appearance of another Li ordering phenomenon.

The CoO$_2$ bulk sample was shiny gray in color and easily collapsed into fine powder. In Fig. 2, XRPD patterns for both end members, pristine LiCoO$_2$ ($x$ = 1.0) and CoO$_2$ ($x$ = 0.0), are shown. For LiCoO$_2$ diffraction peaks were readily indexed based on space group *R*-3*m* with the lattice parameters, $a$ = 2.814 Å and $c$ = 14.05 Å: the $a$ and $c$ values are in good agreement with those previously reported [12,13]. Now the diffraction pattern for CoO$_2$ was indexed based on space group *P*-3*m*1 with the lattice parameters, $a$ = 2.820 Å and $c$ = 4.238 Å. As the use of an airtight sample holder had significantly deteriorated the resolution of diffraction pattern for the CoO$_2$ phase, refinement of the CoO$_2$ crystal structure was unsuccessful. The LiCoO$_2$ phase crystallizes in a so-called "O3-type" structure [17] in which Li ions occupy the octahedral site with three CoO$_2$ layers per unit cell, while the CoO$_2$ phase possesses an "O1-type" structure consisting of a single CoO$_2$ layer only per unit cell (Fig. 3) [12,17]. Accordingly the interlayer distance is much shorter for the CoO$_2$ phase (4.24 Å) than the Li$_x$CoO$_2$ phase ($c$/3 = 4.67 – 4.82 Å). Thus, as $x$ reaches 0 an abrupt shrinkage of the interlayer distance occurs to cause some mechanical fragility of the CoO$_2$ bulk sample.

The key factor for successful synthesis of a bulk sample of CoO$_2$ without additives is to minimize the overpotential at the final stage of electrochemical oxidation procedure. Amatucci *et al*. demonstrated [12] that the CoO$_2$ phase forms at an extremely high voltage, *i.e.* $V$ = +5.2 V for their Li$_x$CoO$_2$/Li cell. Since this potential value is close to the upper limit of the stability window for both the cell material and the anhydrous electrolytes, too high overpotential would induce decomposition/corrosion reactions inside the electrochemical cell. Utilization of an aluminum anode in the present work is considered to have stabilized the electrochemical cell, for the less ionization tendency of aluminum than lithium. It was also found that moderately densified bulks sintered at 900°C in $O_2$ flow are most appropriate for the present electrochemical oxidation procedure. In fact, optimization of the grain microstructure is crucially important to suppress the overpotential. Preliminary results have revealed that the electrochemical de-intercalation of lithium from densified or porous bulks results in an instability of the cell voltage. This suggests that some side reactions inevitably occur in this electrochemical procedure under such conditions.

Shown in Fig. 4 are a thermogravimetric (TG) curve and the corresponding differential TG (DTG) curve taken for a $CoO_2$ sample in 5% $H_2$/Ar gas flow. In this experiment, approximately 3.3 mg of the $CoO_2$ sample was heated with a heating rate of +2°C/min up to 600°C, followed by a rapid cooling to room temperature. Three distinct peaks and steps are seen in the DTG and TG curves below 500°C. The first DTG peak at ~100°C may be attributed to the vaporization of residual electrolyte. The second and third peaks correspond to the following reductive reactions, respectively:

$$CoO_{2-\delta} \rightarrow CoO + (1-\delta)/2\ O_2 \quad (1)$$

$$CoO \rightarrow Co + 1/2\ O_2 \quad (2)$$

For Reactions (1) and (2), both of which accompany weight losses, the amounts of oxygen evolved during individual reactions are obtained to be 0.98 and 1.00, respectively. Thus, the value of $(2-\delta)$ of the starting material is determined at 1.98±0.02. Thus the resultant $CoO_2$ sample is essentially oxygen stoichiometric. Previously, Tarascon *et al.* [17] reported based on their synchrotron x-ray diffraction analysis that their "$CoO_2$" sample prepared by electrochemical de-intercalation of lithium contained both oxygen-stoichiometric $CoO_2$ and oxygen-deficient $CoO_{1.92}$ phases. On the other hand, Venkatraman *et al.* [17] reported a large amount of oxygen deficiency (2–δ = 1.72 – 1.88) for their "CoO$_2$" sample obtained with a chemical oxidation technique. Therefore, it is considered that the oxygen content of CoO$_2$ samples depends strongly on the synthesis procedure such that the more equilibrated synthesis and/or more highly oxidizing conditions tend to suppress the formation of oxygen vacancies.

Magnetic susceptibility ($\chi$) data for the resultant CoO$_2$ sample (solid circles) and the starting LiCoO$_2$ sample (open triangles) are given in Fig. 5. The $\chi$ value for the LiCoO$_2$ sample is small in magnitude and little dependent on temperature, as the constituent Co$^{III}$ is in non-magnetic low-spin state. This result is in good agreement with those previously reported [19]. Surprisingly, the CoO$_2$ sample also exhibits temperature-independent susceptibility in a temperature range between 50 and 300 K. This is in sharp contrast to that for Na$_x$CoO$_2$ with $x > 0.5$ where a Curie-Weiss behavior is clearly observed [4,20]. The $\chi$ *vs T* plots were fitted with the following formula:

$$\chi = \chi_0 + C/(T - \Theta), \qquad (3)$$

where $\chi_0$, $C$, and $\Theta$ denote a constant susceptibility, the Curie constant, and the Weiss temperature, respectively. The values of $\chi_0$, $C$, and $\Theta$ for the CoO$_2$ (LiCoO$_2$) sample were determined to be: 5.68×10$^{-4}$ (1.34×10$^{-4}$) emu/mol Oe, 4.01×10$^{-3}$ (8.87×10$^{-4}$) emu K/mol Oe, and −3.3 (−1.6) K, respectively. The $C$ value for the CoO$_2$ sample yields an effective magnetic moment ($\mu_{eff}$) of 0.18 $\mu_B$/Co site. This value is much smaller than the theoretical "spin-only" value of low-spin Co$^{IV}$ ($S = 1/2$), *i.e.* 1.73 $\mu_B$/Co site. This indicates that any localized spin model is unlikely. Taking into account this low value of $\mu_{eff}$ and also the small value of $\Theta$, the upturn behavior at low temperatures should be considered to be due to an extrinsic cause, *e.g.* lattice defects.

Now we conclude that the magnetism of $CoO_2$ is featured with a temperature-independent susceptibility with a relatively large value for $\chi_0$. A possible (and the most conservative) explanation for this is that the $CoO_2$ phase is a Pauli-paramagnetic metal with itinerant electrons. Assuming that the difference in magnitude of $\chi_0$ between the $CoO_2$ and $LiCoO_2$ phases, *i.e.* $\Delta\chi_0 = 4.3\times10^{-4}$ emu/mol Oe, is the Pauli-paramagnetic contribution [21], the density of states at the Fermi level, $D(\varepsilon_F)$, is calculated at 13 electrons / eV. This value is three times larger than the theoretical value for $D(\varepsilon_F)$, *i.e.* ~4 electrons / eV for non-magnetic $CoO_2$ according to LDA calculations [9]. This may be attributed to an electron-mass enhancement. We note that some unconventional magnetic models such as a spin liquid picture in a frustrated spin lattice [22] may not be ruled out only with the present dc-magnetic susceptibility data. Further detailed characterizations are highly desirable to gain the deeper insight into the electronic structure of the $CoO_2$ phase by means of , *e.g.* $^{59}$Co NMR.

4. Conclusions

Single-phase bulk samples of $CoO_2$ were successfully obtained through electrochemical de-intercalation of lithium from a pristine $LiCoO_2$ sample. The $CoO_2$ phase was found to possess a crystal structure consisting of triangular lattice layers of $CoO_2$ stacked infinitely. Thermogravimetric analysis revealed no significant oxygen deficiency such that the $CoO_2$ sample may be concluded to be essentially oxygen stoichiometric. Magnetic susceptibility of the $CoO_2$ phase little depends on temperature.

The magnitude of initial susceptibility ($\chi_0$) was relatively high. This suggests that the $CoO_2$ phase is a Pauli-magnetic metal with itinerant electrons.

The authors thank Mr. S. Nakamura of Center for Advanced Materials Analysis Technical Department, Tokyo Institute of Technology, for ICP-AES analysis. The present work was supported by Grants-in-aid for Scientific Research (Contract Nos. 16740194 & 19740201) from the Japan Society for the Promotion of Science. M. K. acknowledges financial support from the Academy of Finland (Decision No. 110433).

Figure captions

Fig. 1:

(a) Change in voltage for the $Li_xCoO_2$/Al electrochemical cell as lithium is de-intercalated from $LiCoO_2$. The lithium content ($x$) of $Li_xCoO_2$ is calculated based on Faraday's law. (b) The derivative, $dV/dx$, is plotted in terms of $x$ to elucidate more clearly the de-intercalation process.

Fig. 2:

X-ray powder diffraction patterns for $CoO_2$ (upper) and pristine $LiCoO_2$ (bottom) samples.

Fig. 3:

Crystal structures of pristine $LiCoO_2$ (left) and $CoO_2$ (right) phases. Note that $CoO_2$ crystallizes in a simple "O1-type" structure consisting of only a single $CoO_2$ layer per unit cell, while $LiCoO_2$ forms a rock-salt related "O3-type".

Fig. 4:

Thermogravimetric (TG) curve for $CoO_2$ sample kept in 5% $H_2$/Ar gas flow (lower). The upper panel shows the corresponding differential TG (DTG) curve, revealing two major steps for the compound to decompose completely to Co metal.

Fig. 5:

Temperature dependence of magnetic susceptibility ($\chi$) for the samples of $CoO_2$ (solid circles) and pristine $LiCoO_2$ (open triangles).

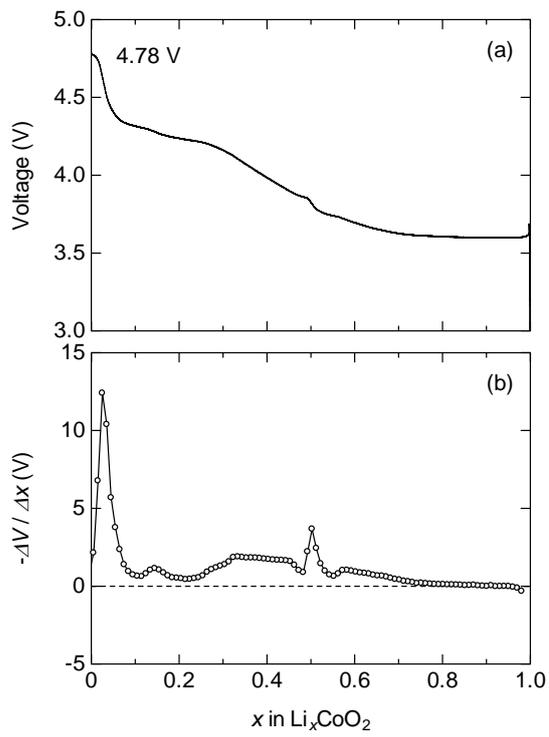

Fig. 1. Motohashi *et al.*

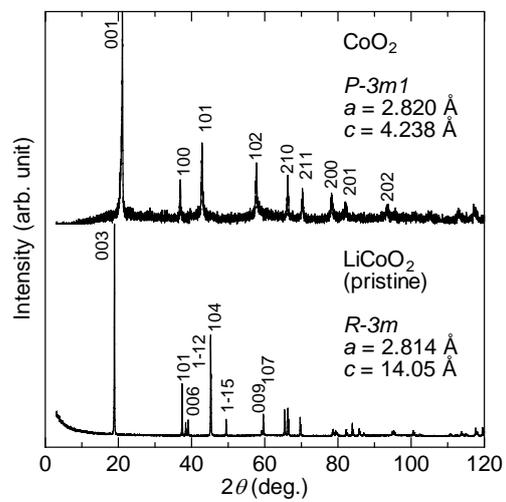

Fig. 2. Motohashi *et al.*

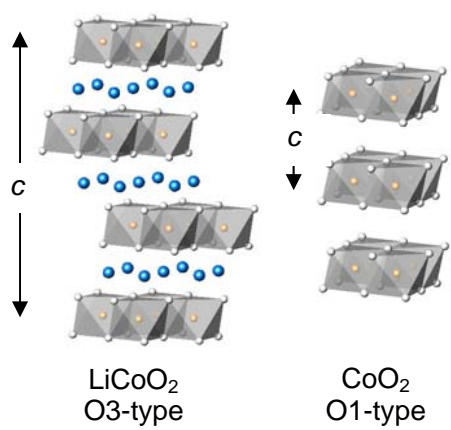

LiCoO$_2$
O3-type

CoO$_2$
O1-type

Fig. 3. Motohashi *et al.*

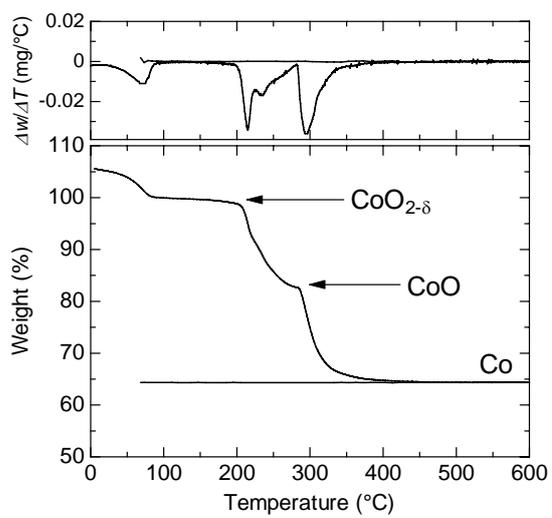

Fig. 4. Motohashi *et al.*

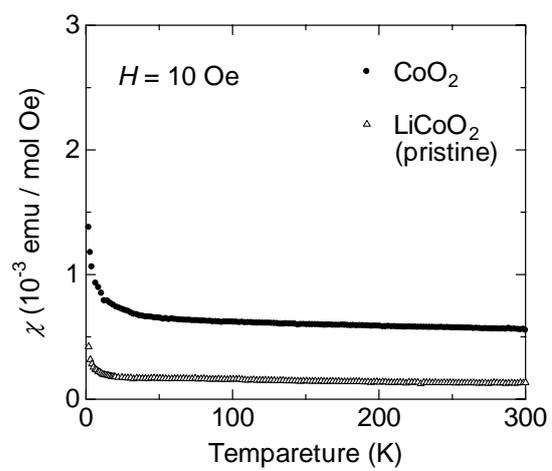

Fig. 5. Motohashi *et al.*